\documentclass{PoS}

\newcommand{\beq}{\begin{equation}}
\newcommand{\eeq}{\end{equation}}

\title{Simulating Wilson fermions \\
without critical slowing down}

\ShortTitle{Wilson fermions without critical slowing down}

\author{\speaker{Urs Wenger}\\
        Albert Einstein Center for Fundamental Physics \\
        Institute for Theoretical Physics\\
        University of Bern,
        Sidlerstr.~5, CH-3012 Bern, Switzerland\\
        E-mail: \email{wenger@itp.unibe.ch}}

      \abstract{ 
        We present a simulation algorithm for Wilson fermions based
        on the exact hopping expansion of the fermion action. The algorithm
        essentially eliminates critical slowing down by sampling the fermionic
        two-point correlation function and it allows simulations directly in
        the massless limit. As illustrative examples, the algorithm is applied
        to the Gross-Neveu and the Schwinger model, the latter in the strong
        coupling limit.  
      }

\FullConference{The XXVII International Symposium on Lattice Field Theory - LAT2009\\
		 July 26-31 2009\\
		 Peking University, Beijing, China}

\begin{document}

\section{Introduction}
Simulating strongly interacting fermions continues to be a major challenge in
computational physics.  The standard procedure to deal with fermionic degrees
of freedom is to integrate out the fermionic fields in order to obtain the
fermion determinant $\det \, D$, where $D$ denotes the Dirac operator.
However, this procedure is not unproblematic. Consider for example a fermion
interacting with a bosonic field $U$.  After integrating out the fermion
fields one obtains $\det\, D(U)$ which yields an effective action non-local in
the bosonic field. The standard method is now to re-express the determinant
using bosonic 'pseudo-fermions' and use the Hybrid Monte Carlo algorithm
\cite{Duane:1987de} which in essence encodes the non-locality of the fermion
determinant in the inverse $D(U)^{-1}$.  Another problem is that the standard
approach suffers from critical slowing down (CSD) towards the chiral limit. In
that limit the correlation length of the fermionic two-point function
diverges. As a consequence the Dirac operator $D(U)$ develops very small modes
and eventually the inverse $D(U)^{-1}$ becomes ill-conditioned.  Yet another
problem concerns the phase of $\det D$ which for Wilson fermions is in general
non-zero.  Hence a probabilistic interpretation of the integration measure,
necessary for any Monte Carlo simulation, is not possible and leads to a sign
problem when an odd number of Wilson
fermion flavours is simulated.\\
\indent Here we propose a novel approach \cite{Wenger:2008tq} circumventing
the above mentioned problems. It is based on the exact hopping expansion of
the fermion action, i.e.~a reformulation of the fermion system as a
statistical closed loop model.  We develop a simulation algorithm which
samples directly the fermionic two-point function and in this way eliminates
CSD. Moreover, it allows to specify the fermionic boundary conditions a
posteriori, i.e.~after the simulation, and allows simulations directly in the
massless limit.  The approach is applicable to the Gross-Neveu (GN) model in
$D=2$ dimensions, to the Schwinger model in the strong coupling limit in $D=2$
and $D=3$ dimensions, to supersymmetric quantum mechanics and the $N=1$ and 2
supersymmetric Wess-Zumino model in $D=2$ dimensions.  In the present
proceedings we concentrate on the application to
the GN and the Schwinger model.\\
\indent Finally, we would like to emphasise that the reformulation based on
the hopping expansion is not new
\cite{Karowski:1984ih,Montvay:1989kj,Gattringer:1998ri,Gattringer:1998cd}.
Mostly, however, it has been applied to staggered fermions in the strong
coupling limit where a reformulation in terms of monomers and dimers
\cite{Rossi:1984cv} allows efficient algorithms
\cite{Karsch:1988zx,Adams:2003cca} that were subsequently applied to many
interesting systems
\cite{Chandrasekharan:2003im,Chandrasekharan:2006tz,deForcrand:2009dh}, see
also the recent review by Chandrasekharan \cite{Chandrasekharan:2008gp}.  For
Wilson fermions on the other hand the loop formulation has been developed for
the Schwinger model in the strong coupling limit \cite{Salmhofer:1991cc} and
the GN model \cite{Gattringer:1998ri,Gattringer:1998cd,Wolff:2007ip} and what
we propose in \cite{Wenger:2008tq} is just a very efficient algorithm for
these loop formulations.

\section{Loop formulation of Wilson fermions}
We start with the reformulation of $D=2$ fermionic systems involving Wilson
fermions in terms of a statistical loop gas model.  We use the GN model, a
prototype for strongly interacting fermions, as an illustrative example.  The
model is most naturally formulated in terms of Majorana fermions.  Employing
the Wilson lattice discretisation for a Majorana fermion the Euclidean
Lagrange density reads
\beq
 {\cal L} = \frac{1}{2} \xi^T {\cal C} (\gamma_\mu \tilde \partial_\mu
 - \frac{1}{2} \partial^* \partial + m) \xi - 
  \frac{g^2}{4} \left( \xi^T {\cal C} \xi \right)^2
\eeq
where $\xi$ is a real 2-component Grassmann field, ${\cal C} = -{\cal C}^T$ is
the charge conjugation matrix and $\partial, \partial^*$ and $\tilde \partial$
are the forward, backward and the symmetric lattice derivative, respectively.
In the continuum, the massless model enjoys a discrete chiral symmetry $\xi
\rightarrow \gamma_5 \xi$ which on the lattice is broken explicitly by the
Wilson term $\frac{1}{2} \partial^* \partial$. The symmetry can be restored in
the continuum by fine tuning $m\rightarrow m_c$. Further we note that a pair
of Majorana fermions may be considered as one Dirac fermion, i.e.~$\psi =
1/\sqrt{2}(\xi_1 + i \xi_2), \quad \overline \psi = 1/\sqrt{2}(\xi_1^T - i
\xi_2^T) {\cal C}$, exposing the $O(2N)$ flavour symmetry explicitly. Since
integrating out Majorana fermions yields the Pfaffian of the antisymmetric
Dirac operator, the model with $2N$ Majorana fermions is equivalent to $N$
Dirac fermions through the identity $ \left(\textrm{Pf} \,D \right)^{2N} =
\left(\det D \right)^N$.

At non-vanishing coupling $g \neq 0$ one usually employs a
Hubbard-Stratonovich transformation and introduces the scalar field $\sigma
\propto \xi^T {\cal C} \xi$. With $M(x) = 2 + m +\sigma(x)$ and $P(\pm \mu) =
\frac{1}{2}(1\mp\gamma_\mu)$ the action then becomes the sum of monomer and
hopping terms
\beq
 S_\textrm{\tiny GN} = \frac{1}{2} \sum_x \xi^T(x) {\cal C} M(x) \xi(x) - \sum_{x,\mu}  \xi^T(x) {\cal C} P(\mu) \xi(x+\hat\mu) .
\eeq
Using the nil-potency of Grassmann elements one can now expand the Boltzmann
factor and perform an exact hopping expansion for the Majorana Wilson fermions
\cite{Wolff:2007ip}. We emphasise that this can be done for any fermionic
theory (bilinear in the fermionic fields). At each site, the fields $\xi^T
{\cal C}$ and $\xi$ must be exactly paired in order to give a non-vanishing
contribution to the path integral,
\beq
\int {\cal D}\xi \, \prod_x \left(M(x)/2 \, \xi^T(x)
{\cal C}  \xi(x) \right)^{m(x)} \prod_{x,\mu}\left(\xi^T(x) {\cal C} P(\mu) \xi(x+\hat \mu)\right)^{b_\mu(x)}
\eeq
where the occupation numbers $m(x) = 0,1$ for monomers and $b_\mu(x) = 0,1$
for bonds (or dimers) satisfy the constraint
\beq
m(x) + \frac{1}{2}\sum_\mu b_\mu(x) = 1.
\eeq
This constraint encodes that only closed, non-intersecting paths survive the
integration and we end up with a closed loop representation of the partition
function in terms of monomers and dimers, i.e.~$Z = \sum_\ell \omega(\ell)$.
The weight $\omega$ of each loop $\ell$ can be calculated analytically
\cite{Gattringer:1998ri,Gattringer:1998cd,Wolff:2007ip,Stamatescu:1980br}
yielding $|\omega(\ell)| = 2^{-c/2}$ where $c$ is the number of corners in the
loop, while the phase of $\omega(\ell)$ depends on the geometrical shape of
$\ell$. In $D=2$ dimensions and for a torus geometry of the lattice,
$\textrm{sign}[\omega(\ell)] \in \{-1,1\}$ depends on the boundary conditions
(BC) $\epsilon_\mu \in \{0,1\}$ and on the number $n_\mu$ of loop windings in
direction $\mu$,
\beq
\textrm{sign}[\omega(\ell)] = (-1)^{n_\mu  (\epsilon_\mu + n_\mu)} \, .
\eeq
As a consequence the overall sign of a given configuration depends only on the
fermionic BC and the total winding number $l=\{l_\mu\}$ (modulo 2).

If we separate all configurations into the equivalence classes ${\cal L}_{ij}$
where the subscripts $i,j$ specify the total winding numbers $l_\mu$ (modulo
2) in the two directions, then the partition function summing over all
non-oriented, self-avoiding loops with positive weight,
\beq
Z =  \sum_{\{\ell\}\in{\cal L}} |\omega[\ell]| \prod_{x \notin \ell} M(x),\quad
{\cal L} \in {\cal L}_{00} \cup {\cal L}_{10} \cup {\cal L}_{01} \cup {\cal
  L}_{11} \,,
\eeq
represents a system with unspecified fermionic BC while systems with specific
fermionic BC can be constructed a posteriori by taking the signs of each class according to
\beq
Z^{\epsilon}_\xi = 2 Z_{{\cal L}_{00}} - \sum_{i,j=0}^{1} (-1)^{\epsilon_\mu l_\mu}
Z_{{\cal L}_{ij}} \,.
\label{eq:fermionic partition function}
\eeq
Finally we note that if one considers $N>1$ Majorana flavours the occupation
numbers $m, b_\mu$ are decorated by the flavour index $\alpha$ and one
considers $N$ different loop flavours. The monomer weight $M(x)$ depends on
the local fermion density $\sum_\alpha m^\alpha(x)$ only and one ends up with a
model of locally coupled loops.

In the Schwinger model the hopping term contains a $U(1)$ phase coming from
the gauge field $\phi_\mu (x)$, and the non-oriented (Majorana) bonds carry an
additional factor $\propto \cosh(\phi_\mu(x))$. Moreover the gauge field
introduces an interaction between the two Majorana flavours proportional to
$\pm \sinh(\phi_\mu(x))$, These additional factors introduce a sign problem
since each loop can now have an arbitrary sign.  However, in the strong
coupling limit, the two flavours are bound together. In the present
formulation it means that two different Majorana loops lay on top of each
other and the resulting double loop describes the world line of the bosonic
bound state. It also turns out that all the signs cancel in a non-trivial way
and so the bosonisation is realised explicitly. Eventually we end up with a
model of non-oriented loops \cite{Salmhofer:1991cc} in which all the loop and
monomer weights are squared compared to the GN model. Note further that
eq.(\ref{eq:fermionic partition function}) no longer applies because the
fermionic BC have no impact on the BC of the corresponding bosonic bound state
-- instead the relevant partition function is the one where all topological
classes contribute positively, i.e.~$Z$.
\begin{figure}[t]
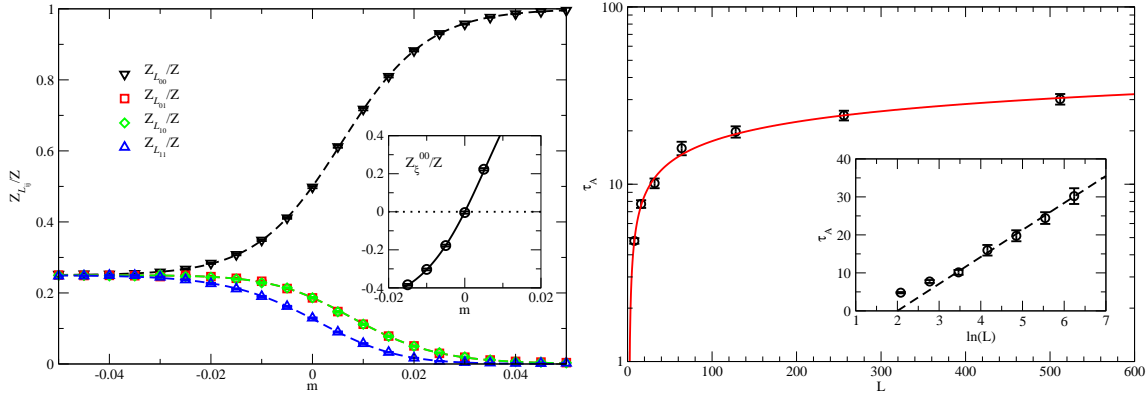

\includegraphics[width=7.5cm]{Figures/gn_partition_function_ratios_L128T128_nCP20000x100_2} \hfill
\includegraphics[width=7.5cm]{Figures/integrated_autocorrelation_time_gn_alt_new}
\caption{$N=1$ Majorana GN model on a $128^2$ lattice. Left: Comparison of
  simulation results (symbols) and analytic calculations (dashed lines) for the
  partition function ratios $Z_{{\cal L}_{ij}}/Z$. The inset shows the
  reproduction of the zero mode of $Z_\xi^{00}$ at $m_c=0$. Right: Integrated
  autocorrelation time of the condensate at the critical point $m_c=0$ fitted by $\tau_A \sim L^z$
  with $z=0.31(4)$. The inset shows a fit to a logarithmic dependence on $L$.
\label{fig:gn_partition_function_ratios_L128T128_nCP20000x100_2}
}
\end{figure}

\section{Simulation algorithm for loops and strings}
A standard procedure to simulate loop gas models as the one described above is
to perform local loop updates involving plaquette moves only
\cite{Gausterer:1992ud,Gattringer:2007em}. One problem with such an algorithm
is that it can not change between the topological classes ${\cal L}_{00},{\cal
  L}_{10},{\cal L}_{01},{\cal L}_{11}$.  Moreover, if the correlation length
of the system grows large these algorithms become highly inefficient and
suffer from CSD.  Our proposal \cite{Wenger:2008tq} (subsequently worked out
in \cite{Wolff:2008xa}) follows the one of Prokof'ev and Svistunov
\cite{Prokof'ev:2001zz} and enlarges the configuration space by open fermionic
strings.  In the GN model an open string corresponds to the insertion of a
Majorana fermion pair $\{\xi(x),\xi^T(y) {\cal C}\}$ at position $x$ and $y$
into the path integral, and the open string samples directly the correlation
function
\beq
G(x,y) = \int {\cal D} \xi e^{-S_\textrm{\tiny GN}} \xi(x) \xi(y)^T {\cal C}\, .
\eeq
This is the reason why CSD is eliminated: configurations are updated on all
length scales up to $O(\zeta)$ where $\zeta$ is the correlation length
corresponding to the fermionic two point function. As a consequence the update
remains efficient even at a critical point where the correlation length
diverges.  Contact with the partition functions $Z_{{\cal L}_{ij}}$ is made
each time the open string closes and this provides the proper normalisation
for the expectation value of the 2-pt.~function, $\langle \xi(x) \xi(y)^T
{\cal C} \rangle_Z = G(x,y)/Z$, or any other observables.  In practice, the
ends of the open string are updated with a standard local Metropolis or heat
bath procedure \cite{Wenger:2008tq}. Similar ideas have been around for a long
time in various other contexts
\cite{Prokof'ev:2001zz,Evertz:1992rb,Syljuasen:2002zz} -- what is new here is
the practical application to Wilson fermions and the demonstration that CSD is
essentially eliminated.

\section{Absence of critical slowing down}
\begin{figure}[t]
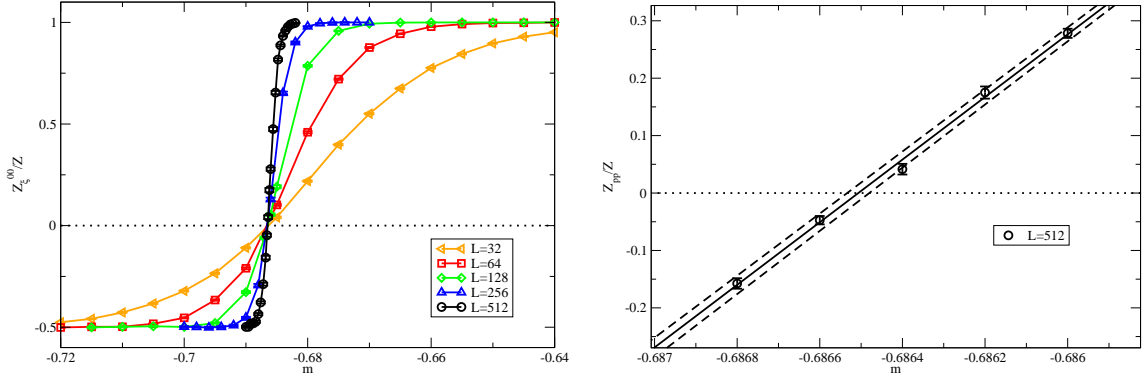

\includegraphics[width=7.5cm]{Figures/zpp_original} \hfill
\includegraphics[width=7.2cm]{Figures/zpp_original_zoom}
\caption{The Schwinger model in the strong coupling limit. Left: Partition
  function ratio $Z_\xi^{00}/Z$ on various lattices. Right: Determination of
  the critical point $m_c = -0.686506(27)$ on the largest lattice with $L=512$.
\label{fig:SCSM}
}
\end{figure}
Before investigating the efficiency of the algorithm, we demonstrate its
correctness by comparing simulation results with analytically
know expressions. For this purpose we use the $N=1$ Majorana GN model. This
model is essentially a free fermion model and can be solved exactly by
calculating Pfaffians in momentum space.
In the left plot of Figure
\ref{fig:gn_partition_function_ratios_L128T128_nCP20000x100_2} we show the
results for the partition function ratios $Z_{{\cal L}_{ij}}/Z$ on a $128^2$
lattice from 2M closed path configurations (symbols) compared to the exact
results (dashed lines).  The inset shows the combination $Z_\xi^{00} =
Z_{{\cal L}_{00}} - Z_{{\cal L}_{10}} - Z_{{\cal L}_{01}} - Z_{{\cal L}_{11}}$
which has a zero mode at the critical point $m_c=0$. The algorithm is indeed
able to reproduce the zero mode without problems. In order to investigate the
efficiency of the algorithm at the critical point we measure the condensate
$\langle \xi^T {\cal C} \xi\rangle_{Z_\xi}$.  The right plot of Figure
\ref{fig:gn_partition_function_ratios_L128T128_nCP20000x100_2} shows the
integrated autocorrelation time $\tau_A$ of the condensate as a function of
the linear system size $L$. The dynamic exponent $z$ relevant for CSD,
i.e.~$\tau_A \sim L^z$, turns out to be $z \simeq 0.31(4)$. A dependence
logarithmically on $L$ can also be fitted to $L \geq 32$ yielding
$-14.2(2.5)+7.1(6) \ln(L)$ with $\chi^2/\textrm{dof}=0.18$.

Next we consider the Schwinger model in the strong coupling limit
$g\rightarrow \infty$ as a non-trivial example for strongly interacting
fermions. In the left plot of Figure \ref{fig:SCSM} we show the
partition function ratio $Z_\xi^{00}/Z$ on various lattices up to $L=512$. As
in the Majorana GN model we find a zero of the partition function which
depends only very little on the extent of the lattice. We can use
$Z_\xi^{00}(m_c) = 0$ as a definition for the critical point $m_c$. It can be
determined by a linear fit and we obtain $m_c=-0.686506(27)$ (cf.~right plot
in Figure \ref{fig:SCSM}) from our simulations on the largest lattice with
$L=512$. Further improvement could be achieved by employing standard
reweighting techniques as done in \cite{Gausterer:1995jh} where they obtained
$m_c=-0.6859(4)$. These calculations indicated a second order phase transition
in the universality class of the Ising model (with critical exponent
$\nu\simeq 1$). 
\begin{figure}[t]
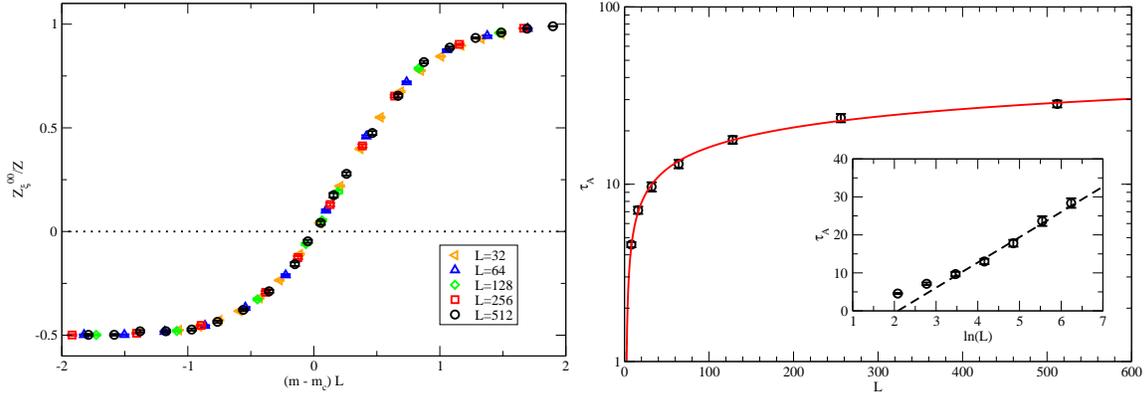

\includegraphics[width=7.5cm]{Figures/zpp_FSscaling_original} \hfill
\includegraphics[width=7.5cm]{Figures/integrated_autocorrelation_time_scsm_alt}
\caption{The Schwinger model in the strong coupling limit. Left: Finite size
  scaling of $Z_\xi^{00}/Z$ for a second order phase transition in the
  universality class of the Ising model. Right: Integrated autocorrelation time
  of the condensate at the critical point $m_c$ fitted by $\tau_A \sim L^z$
  with $z=0.25(2)$. The inset shows a fit to a logarithmic dependence on $L$.
\label{fig:SCSM2}
}
\end{figure}
Our results in the left plot of Figure \ref{fig:SCSM2} now confirm this by
demonstrating that the partition function ratios $Z_\xi^{00}/Z$ as a function
of the rescaled mass $(m - m_c) L^\nu$ with $\nu=1$ beautifully collapse onto
a universal scaling curve. The efficiency of the algorithm and the fact that
CSD is essentially absent is demonstrated in the right plot of
Fig.~\ref{fig:SCSM2} where we show the integrated autocorrelation time
$\tau_A$ of the energy as a function of the linear system size $L$ at the
critical point $m=m_c$.  The functional dependence on $L$ can be well fitted
($\chi^2/\textrm{dof}=1.28)$ by $\tau_A \sim L^z$ all the way down to our
smallest system size $L=8$. We obtain $z=0.25(2)$ which is consistent with
just using the largest two system sizes. The autocorrelation time may also depend logarithmically on
$L$ and a fit to $L \geq 32$ yields $-13.8(1.9) + 6.6(4) \ln(L)$ with
$\chi^2/\textrm{dof}=1.00$. In any case it is an amazing result that our
local Metropolis-type update appears to have a dynamical critical exponent
close to zero.  

\section{Conclusions}
In conclusion, we have presented a new type of algorithm for Wilson fermions
in two dimensions. It relies on sampling directly 2-point correlation
functions and essentially eliminates critical slowing down. We have
successfully tested our algorithm on the Majorana GN model and on the
Schwinger model in the strong coupling limit and found remarkably small
dynamical critical exponents.  The algorithm definitely opens the way to
simulate efficiently generic loop models (with positive weights) in arbitrary
dimensions, in particular the GN model with any number of flavours, the
Thirring model, the Schwinger model and QED$_3$ in the strong coupling limit,
as well as fermionic models with Yukawa-type scalar interactions like the
$N=1$ and 2 Wess-Zumino models, all with Wilson fermions.


\begin{thebibliography}{10}

\bibitem{Duane:1987de}
S.~Duane, A.~D. Kennedy, B.~J. Pendleton and D.~Roweth,
\newblock Phys. Lett. {\bf B195}, 216 (1987).

\bibitem{Wenger:2008tq}
U.~Wenger,
\newblock Phys. Rev. {\bf D80}, 071503 (2009), [arXiv:0812.3565].

\bibitem{Karowski:1984ih}
M.~Karowski, R.~Schrader and H.~J. Thun,
\newblock Commun. Math. Phys. {\bf 97}, 5 (1985).

\bibitem{Montvay:1989kj}
I.~Montvay,
\newblock Phys. Lett. {\bf B227}, 260 (1989).

\bibitem{Gattringer:1998ri}
C.~Gattringer,
\newblock Int. J. Mod. Phys. {\bf A14}, 4853 (1999), [cond-mat/9811139].

\bibitem{Gattringer:1998cd}
C.~Gattringer,
\newblock Nucl. Phys. {\bf B543}, 533 (1999), [hep-lat/9811014].

\bibitem{Rossi:1984cv}
P.~Rossi and U.~Wolff,
\newblock Nucl. Phys. {\bf B248}, 105 (1984).

\bibitem{Karsch:1988zx}
F.~Karsch and K.~H. Mutter,
\newblock Nucl. Phys. {\bf B313}, 541 (1989).

\bibitem{Adams:2003cca}
D.~H. Adams and S.~Chandrasekharan,
\newblock Nucl. Phys. {\bf B662}, 220 (2003), [hep-lat/0303003].

\bibitem{Chandrasekharan:2003im}
S.~Chandrasekharan and F.-J. Jiang,
\newblock Phys. Rev. {\bf D68}, 091501 (2003), [hep-lat/0309025].

\bibitem{Chandrasekharan:2006tz}
S.~Chandrasekharan and F.-J. Jiang,
\newblock Phys. Rev. {\bf D74}, 014506 (2006), [hep-lat/0602031].

\bibitem{deForcrand:2009dh}
Ph.~de~Forcrand and M.~Fromm,
\newblock arXiv:0907.1915.

\bibitem{Chandrasekharan:2008gp}
S.~Chandrasekharan,
\newblock PoS {\bf LATTICE2008}, 003 (2008), [arXiv:0810.2419].

\bibitem{Salmhofer:1991cc}
M.~Salmhofer,
\newblock Nucl. Phys. {\bf B362}, 641 (1991).

\bibitem{Wolff:2007ip}
U.~Wolff,
\newblock Nucl. Phys. {\bf B789}, 258 (2008), [arXiv:0707.2872].

\bibitem{Stamatescu:1980br}
I.~O. Stamatescu,
\newblock Phys. Rev. {\bf D25}, 1130 (1982).

\bibitem{Gausterer:1992ud}
H.~Gausterer, C.~B. Lang and M.~Salmhofer,
\newblock Nucl. Phys. {\bf B388}, 275 (1992).

\bibitem{Gattringer:2007em}
C.~Gattringer, V.~Hermann and M.~Limmer,
\newblock Phys. Rev. {\bf D76}, 014503 (2007), [arXiv:0704.2277].

\bibitem{Wolff:2008xa}
U.~Wolff,
\newblock Nucl. Phys. {\bf B814}, 549 (2009), [arXiv:0812.0677].

\bibitem{Prokof'ev:2001zz}
N.~Prokof'ev and B.~Svistunov,
\newblock Phys. Rev. Lett. {\bf 87}, 160601 (2001).

\bibitem{Evertz:1992rb}
H.~G. Evertz, G.~Lana and M.~Marcu,
\newblock Phys. Rev. Lett. {\bf 70}, 875 (1993), [cond-mat/9211006].

\bibitem{Syljuasen:2002zz}
O.~F. Syljuasen and A.~W. Sandvik,
\newblock Phys. Rev. {\bf E66}, 046701 (2002).

\bibitem{Gausterer:1995jh}
H.~Gausterer and C.~B. Lang,
\newblock Nucl. Phys. {\bf B455}, 785 (1995), [hep-lat/9506028].

\end{thebibliography}
\end{document}